\documentstyle[12pt]{article}
\makeatletter
  
  \@addtoreset{equation}{section}
\makeatother

\def\abstract#1{\vskip 7mm 
	\begin{center}{\large Abstract}\par \bigskip
		\begin{minipage}[c]{12cm}
			\small #1
		\end{minipage}
	\end{center}
}
\def\title#1{\begin{center}{\Large\bf #1}\end{center}}
\def\author#1{\vskip 5mm \begin{center}{#1}\end{center}}
\def\address#1{\begin{center}{\it #1}\end{center}}

\newcommand{\bfr}{\begin{flushright}}
\newcommand{\efr}{\end{flushright}}

\begin{document}

\vspace*{-2cm}
\bfr{}\efr\vspace{-9mm}
\bfr{}\efr
\vspace{0.5cm}

\title{Moyal Quantization for\\ Constrained System}
\author{Takayuki HORI\footnote{E-mail: hori@main.teikyo-u.ac.jp},~Takao KOIKAWA\footnote{E-mail: koikawa@otsuma.ac.jp} and Takuya MAKI\footnote{E-mail: vk7t-mk@asahi-net.or.jp}}
\vspace{1cm}
\address{
	$^{*)}$ Department of Economics, Teikyo University,\\
	Hachioji 192-0395,~Janan\\
        $^{\dagger)}$ School of Social Information Studies,\\
        Otsuma Women's University\\
        Tama 206-0035,~Japan\\ 
        $^{\ddagger)}$ Japan Women's College of Physical Education,\\
        Setagaya,~Tokyo 157-8565,~Japan\\
}
\vspace{1cm}

We study the Moyal quantization for the constrained system. One of the purposes is to give a proper definition of the Wigner-Weyl(WW) correspondence, which connects the Weyl symbols with the corresponding quantum operators. A Hamiltonian in terms of the Weyl symbols  becomes different from the classical Hamiltonian for the constrained system, which is related to the fact that the naively constructed WW correspondence is not one-to-one any more. In the Moyal quantization a geometrical meaning of the constraints is clear. In our proposal, the 2nd class constraints are incorporated into the definition of the WW correspondence by limiting the phasespace to the hypersurface.  Even though we assume the canonical commutation relations in the formulation, the Moyal brackets between the Weyl symbols yield the same results as those for the constrained system derived by using the Dirac bracket formulation. 

\newpage
\setcounter{page}{2}
\section{Introduction}

The Moyal representation provides an alternative of the quantum mechanics,~\cite{Moy} which is  based on the Wigner-Weyl(WW hereafter) correspondence between operators and phasespace functions called the Weyl symbols. Each quantum observable $\hat A$ is related to a Weyl symbol $A_W(q,p)$ by the WW correspondence.~\cite{OsMo} In most cases, the WW correspondence is a one-to-one map. Therefore, when we know quantum mechanical description by operators, we can immediately find a counterpart of the quantum theory by using the WW correspondence. The procedure of obtaining the Moyal quantization is as follows.  We start with the classical fields, obtain the quantum description of the system by use of the Dirac method, and finally get the Weyl symbol by using the WW correspondence only when it has the  one-to-one map property.
As far as the one-to-one property holds, it is always possible to derive the Weyl symbol from the operator representation by using the WW correspondence, and vice versa.  We can always consider the Moyal quantization in these cases. However this is not always the case. In deriving the WW correspondence, we assume a commutation relation(CR hereafter) between coordinate and momentum operators denoted by $\hat q$ and $\hat p$ in order to find the map from phasespace functions to the operators. The quantum CRs are obtained from the Poisson brackets by replacing $q$ and $p$ by the corresponding operators $\hat q$ and $\hat p$. This is valid for the case where there is no constraint. However, when there exist the 2nd class constraints, it is known that the Poisson brackets are to be replaced by the Dirac brackets.~\cite{Dirac} The quantum CRs obtained from the Dirac brackets are no more the canonical ones for the constrained cases. Then, the Weyl symbols related to these quantum operators are subject to modification by the constraints, since the quantum CRs are changed in the definition of WW correspondence. As the result, we find that the WW correspondence, which is defined in the same way for the non-constrained case, maps the multiple functions to an operator. When there is no constraint, the Weyl symbols and the classical fields are identical because of the one-to-one correspondence. However, when there are some constraints, the one-to-one property of the correspondence breaks and so we can not determine the Weyl symbols uniquely even when we know the quantum theory. This presents a serious problem in the Moyal quantization for the constrained system.

The purpose of the present paper is to study how the WW correspondence should be formulated when there exist constraints. There could be various kinds of constraints that should be quantized following the Dirac's quantization method. We write CRs as
\begin{equation}
[\hat q_i, \hat p_j]=i\hbar g_{ij}(q,p),
\label{eq:gcr}
\end{equation}
where $g_{ij}(q,p)$ are, in general, the functions of $q$ and $p$  and should be determined by the Dirac brackets. They can be put into the form $\delta_{ij}$ for the non-constrained system. There can be a variety of constraints. However, we do not exhaust them, because our goal is to study their effects to the Moyal quantization. In this paper, we limit ourselves to the case where $g_{ij}(q,p)$ is given by a constant matrix which is different from $\delta_{ij}$. Such an example can be found in a constraint that defines the space normal to a constant vector in $R^D$. 

The present paper is organized as follows. In next section we discuss the Moyal quantization for the constrained system, where we adopt the constraint mentioned above. We show two methods of formulating WW correspondence for the constrained case. In the first subsection,  we study the method which is the same as that in the non-constrained case.  The coordinate and momentum operators appear in the exponent of the exponential function in the definition of the WW correspondence. These operators are assumed to satisfy the quantum CRs derived from the Dirac brackets. As the result, we find that the correspondence is no more one-to-one. In the following subsection, we propose a modified WW correspondence where we do not use the exponentiated operators to circumvent the problem. In deriving operators from the functions of the phasespace, the eigen-states instead of the coordinate and momentum operators are introduced in the definition of the correspondence. The observation of the reason why the one-to-one map does not hold for the constrained case leads to a proposal in section 3. We propose a geometrical way of imposing the constraints in the correspondence relating the classical functions to the operators. We assume the canonical CRs between the coordinate and momentum operators. In the simple examples considered here, despite the assumption it turns out to be the same map as that obtained by the WW correspondence with coordinate and momentum operators which do not satisfy the canonical CRs. In other words, we reproduce the same result as that by the Dirac quantization method starting with the canonical CRs between coordinate and momentum operators. Section 4 is devoted to exhibiting the application of the WW correspondence for the constrained system. The first application is the harmonic oscillator constrained to the surface normal to a constant vector. The 2nd is the electro-magnetic dynamics in the Coulomb gauge. The last section is for summary and discussion.

\section{Moyal quantization for the constrained system}

In this section we analyze the Moyal quantization for the constrained system. We study the Moyal quantization when there exists a constraint that defines the flat surface normal to a constant vector. We study two maps. They differ in the point whether we use the coordinate and momentum operators in the definitions of the WW correspondence. In the first formulation in section 2.1, we apply the ordinary WW correspondence, which is used for the non-constrained case, to the constrained case. The exponentiated operators appear in the definition of the WW correspondence. We use the Dirac brackets instead of the Poisson brackets to obtain the quantum CRs which are no more canonical ones. We denote the WW correspondence by the map $\cal O$. By using the map, which yields an operator by acting to a function of the phasespace, we obtain a quantum Hamiltonian when we start with the classical Hamiltonian. However, the Weyl symbol which should be obtained by the inverse map of $\cal O$ from the operator is not determined, because the one-to-one correspondence between the operators and the Weyl symbols breaks in this case. We find that $\cal O$ maps multiple functions to an operator.  The classical Hamiltonian and the Hamiltonian in terms of the Weyl symbols do not agree in the constrained case. In other words, we can not establish the Moyal quantization as far as we use the ordinary WW correspondence.  Therefore, we define another correspondence by using the eigen-states, which respects the one-to-one property. We denote the modified WW correspondence by ${\cal S}$, for which there exists an inverse map ${\cal S}^{-1}$.

\subsection{$\cal O$ formulation of WW correspondence}

First we recapitulate the notations which we use here and in the following sections. We denote the 2D dimensional phasespace coordinates by
\begin{equation}
\zeta = (q_1,q_2,\cdots,q_D,p_1,p_2,\cdots,p_D)=(q,p)\in R^{2D}.
\label{eq:qp}
\end{equation}
The 2D dimensional coordinates dual to these phasespace coordinates are denoted by
\begin{equation}
u=(u_1,u_2)\in R^{2D}.
\end{equation}
We express 2D component operators which is obtained by replacing $q$ and $p$ in (\ref{eq:qp}) by the operators,
\begin{equation}
\hat{z} = (\hat{q}, \hat{p}).
\end{equation}

The Fourier transformation of a function $A(\zeta)$ reads 
\begin{equation}
A(\zeta)=\int du~a(u)e^{iu \cdot \zeta},
\label{eq:invfourier}
\end{equation}
where $a(u)$ is inversely obtained as
\begin{equation}
a(u)=\frac{1}{(2 \pi)^{2D}} \int d\zeta A(\zeta)e^{-iu \cdot \zeta}.
\label{eq:fourier}
\end{equation}
By using this $a(u)$ we define the WW correspondence by replacing $\zeta$ in (\ref{eq:invfourier}) by the operator $\hat z$, 
\begin{equation}
A(\hat z)=\int du~a(u)e^{iu \cdot \hat z}.
\label{eq:oprinvfourier}
\end{equation}
We rewrite this to obtain the expression in a differential form, 
\begin{equation}
\hat A=A(-i\nabla_u)e^{iu \cdot \hat z}|_{u=0},
\label{eq:d_invfourier}
\end{equation} 
where $A(\hat z)$ is $\hat A$. This relation between $\hat A$ and $A(\zeta)$ is the WW correspondence, and we denote it as
\begin{equation}
\hat A={\cal O}(A).
\label{eq:calO}
\end{equation}
This is the replacement of the variable $\zeta$ by the operator $\hat z$ in the Fourier transformation (\ref{eq:invfourier}).

We consider an inverse Fourier transformation for the functional case. Then it is natural to ask how we define the inverse map of ${\cal O}$ in the same way as the ordinary Fourier analysis for functions. 
Actually we can obtain the inverse map explicitly for the non-constrained case where the operators $\hat q$ and $\hat p$ satisfy the canonical CRs. We study how this is modified when there exist constraints, and so the CRs are different from the canonical ones. We assume that the quantum CRs between $\hat q$ and $\hat p$ are given by
\begin{equation}
[\hat q_i, \hat p_j]=i\hbar\Pi_{i,j},
\label{eq:PCR}
\end{equation}
where $\Pi_{ij}$ are singular constant matrix elements with a projection property satisfying $\Pi_{ik}\Pi_{kj}=\Pi_{ij}$. 
Generally $\Pi$ has zero vectors, $n$. 
The explicit example of such a projection will be shown later in the
examples. 
A representation of (\ref{eq:PCR}) is obtained on  functions of $s \in R^D$ as 
\begin{eqnarray} 
\hat q_k &=& \Pi_{kl}s_l,\nonumber \\ 
\hat p_k &=& -i\hbar \Pi_{kl}\frac{\partial}{\partial s_l},\nonumber \\
|x\rangle  &=& \delta^D(s -x).
\end{eqnarray}
The above operators satisfy the CR as well as the constraint equations
\begin{equation}
n \cdot \hat{q} = n \cdot \hat{p} = 0. 
\end{equation}
Either $\hat q$ or $\hat p$ can be represented by using ${\delta}_{ij}$ instead of ${\Pi}_{ij}$ for satisfying CR (\ref{eq:PCR}), though it does not satisfy the constraints. Therefore we exclude these representation. 
The state $|x\rangle$ is the eigenstate of $\hat{q}$: 
\begin{eqnarray} 
\hat q_i|x\rangle=\Pi_{ij}x_j|x\rangle. 
\end{eqnarray}
In order to obtain a functional form from the corresponding operator, note that
\begin{equation}
\langle x|e^{iu \hat z}|y\rangle=e^{iu_1 \cdot \Pi(x+y)/2}\delta(x-y+\hbar \Pi u_2).
\end{equation}
By using this relation, the matrix element of the WW correspondence is given by
\begin{equation}
\langle x|A(\hat z)|y \rangle=h^{-D}\int dp~e^{ip \cdot (x-y)/\hbar}
A_W \left( \frac{\Pi \cdot(x+y)}{2},\Pi p \right).
\end{equation}
By introducing new variables $q$ and $v$ by $q=(x+y)/2$ and $v=x-y$, we obtain
\begin{equation}
A_W(\Pi q,\Pi p)=\int dv~e^{-ip \cdot v/\hbar} \left \langle q+\frac{v}{2} \Biggm |A(\hat z) \Biggm| q-\frac{v}{2} \right \rangle.
\label{eq:matrixfourier}
\end{equation}
We denote this procedure of getting $A_W(\Pi q,\Pi p)$ from $\hat A=A(\hat z)$ by ${\cal S}^{-1}$:
\begin{equation}
\tilde \Pi A_W={\cal S}^{-1}(\hat A),
\label{eq:c_q}
\end{equation}
where the map $\tilde \Pi$ is defined by
\begin{equation}
\tilde \Pi (A(q,p)) =A(\Pi q,\Pi p).
\end{equation}
We use the notation ${\cal S}^{-1}$ instead of  ${\cal O}^{-1}$, because  ${\cal O}$ is not one-to-one and so the inverse image of ${\cal O}$ does not map to a single function for the constrained case. As far as $\Pi$ is different from $\delta$, 
${\cal S}^{-1}$ does not coincide with the ${\cal O}^{-1}$. 
We show this in the following way. We define $\Delta(q,p)$ by
\begin{equation}
\Delta(q,p)=A(\Pi q,\Pi p)-A(q,p),
\end{equation}
which is not always zero in $R^{2D}$.
Using
\begin{equation}
\int dv~e^{-ip \cdot v/\hbar} \left< q+\frac{v}{2} \Biggm| {\cal O}(\Delta(q,p))
\Biggm| q-\frac{v}{2} \right>=0,
\end{equation}
we obtain
\begin{equation}
{\cal S}^{-1} {\cal O}(\Delta(q,p))=0.
\end{equation}
As we shall discuss in the following subsection, we can construct the inverse of ${\cal S}^{-1}$ explicitly. Therefore this means that ${\cal O}(\Delta(q,p))=0$, which turns out that $A(q,p)+\Delta(q,p)$ and $A(q,p)$ yield the same operator, 
and so the map ${\cal O}$ is not the one-to-one map lest $\Pi_{ij}$ equals
$\delta_{ij}$. This is why we introduce the second notation ${\cal S}^{-1}$.

We next study what the ${\cal S}^{-1}$ map of a product of two operators brings about. This is the $\star$ product between two Weyl symbols for the constrained case. Let $\hat A$ and $\hat B$ be two operators, and consider their product
\begin{equation}
\hat A \hat B=\int du \int dw ~a(u) b(w)e^{iu \cdot \hat z}e^{iw \cdot \hat z}.
\end{equation}
Using the CR in the constrained case, we obtain
\begin{equation}
e^{iu \cdot \hat z}e^{iw \cdot \hat z}=e^{i(u+w) \cdot \hat z}e^{
-\frac{i}{2}\hbar u \cdot J_{\Pi} w},
\end{equation}
where the 2D $\times$ 2D matrix $J_{\Pi}$ is given by
\begin{equation}
J_{\Pi}=\pmatrix{0 & \Pi \cr
           -\Pi & 0\cr
           },
\end{equation}
We have another formula given by
\begin{equation}
\left< q+\frac{\xi}{2} \Biggm| e^{i(u+w) \cdot \hat z} \Biggm| q-\frac{\xi}{2} 
\right>=e^{i(u_1+w_1) \cdot \Pi  q}
\delta(\xi+\hbar \Pi(u_2+w_2)).
\end{equation}
Using these formulas, we compute the ${\cal S}^{-1}$ map of the product of two operators by performing the $\xi$ integration after multiplying $e^{-ip \cdot \xi/\hbar}$
\begin{eqnarray}
&&{\cal S}^{-1}(\hat A \hat B) \nonumber \\
&=&\int d\xi \int du \int dw ~a(u)b(w)
e^{i(u_1+w_1) \cdot \Pi q}
e^{-\frac{i}{2}\hbar u \cdot J_{\Pi} w}
e^{-ip \cdot \xi/\hbar}
\delta(\xi+\hbar \Pi (u_2+w_2)) \nonumber \\
&=&e^{\frac{i}{2}\nabla_z \cdot J_{\Pi} \nabla_{z'}}A(\Pi q,\Pi p)B(\Pi q',\Pi p')|_{z=z'} \nonumber \\
&=&A(\Pi q,\Pi p) \star B(\Pi q,\Pi p) \nonumber\\
&=&\tilde \Pi (A(q,p) \star B(q,p)),
\end{eqnarray}
where the star product of two functions $f(q,p)$ and $g(q,p)$ is given by
\begin{equation}
f(q,p)\star g(q,p)={\rm exp}\Big\{\frac{i \hbar}{2} \Pi_{ij}
\Big(\frac{\partial}{\partial q_i}\frac{\partial}{\partial p_j'}-
\frac{\partial}{\partial p_i}\frac{\partial}{\partial q_j'}\Big)\Big\}f(q,p)g(q',p')|_{q=q',p=p'}.
\label{eq:starproduct}
\end{equation}
We are able to express the map $\tilde \Pi$ in terms of ${\cal S}^{-1}$ and ${\cal O}$. From Eq.(\ref{eq:calO}) and Eq.(\ref{eq:c_q}), we obtain
\begin{equation}
\tilde \Pi={\cal S}^{-1}{\cal O}.
\end{equation}
By using this relation, above expression reads
\begin{equation}
{\cal S}^{-1}({\cal O}(A) {\cal O} (B))={\cal S}^{-1}{\cal O}(A \star B).
\end{equation}
The existence of the inverse of ${\cal S}^{-1}$ leads to a formula:
\begin{equation}
{\cal O}(A){\cal O} (B)={\cal O}(A \star B).
\label{eq:Ostar}
\end{equation}

By using above formulas we can clarify the correspondence between the operator CR and the functional CR. We define the functional CR by
\begin{equation}
[f(q,p),g(p,p)]=f \star g-g \star f.
\end{equation}
The CRs of operators $\hat{q}_l$ and $\hat{p}_m$ are given by
\begin{equation}
[{\hat{q}}_l, {\hat{p}}_m]_{opr}=i \hbar \Pi_{lm}.
\label{eq:oprcr}
\end{equation}
By using the formula (\ref{eq:Ostar}), their relation is obtained as
\begin{eqnarray}
&&[{\hat{q}}_l, {\hat{p}}_m]_{opr}-i \hbar \Pi_{lm} \nonumber \\
&=&{\cal O} ([q_l,p_m]-i \hbar \Pi_{lm}) \nonumber \\
&=&0.
\end{eqnarray}
Note that ${\cal O}$ is not a one-to-one map in the constrained case because of the multiple-map property of the map. Therefore we can not say that it is an isomorphism. Still, it is possible to regard the  following as a correspondence to (\ref{eq:oprcr})
\begin{equation}
[q_l, p_m]=i \hbar \Pi_{lm}.
\end{equation}

\subsection{${\cal S}$ formulation of WW correspondence}

As is seen in the previous subsection, multiple functions,
 each is different from one another only outside the constraint surface, correspond to a single  
operator by  the ${\cal O}$ maps.
On the other hand ${\cal S}^{-1}$ turns out to be a one-to-one map  
and has its inverse.
It is easy to see the map from the space of functions to the space of operators, defined by
\begin{eqnarray}
          {\cal S}(A) = \int du_2d\zeta~ e^{\frac{i}{\hbar}u_2p}A(\zeta)\biggm|\left.x + \frac{u_2}{2} \right \rangle \left \langle x - \frac{u_2}{2}\right.\biggm| \label{spade}
\end{eqnarray}
is the inverse of ${\cal S}^{-1}$. 
The matrix element of (\ref{spade}) is
\begin{eqnarray}
   \langle x|{\cal S}(A)|x'\rangle  = \int\!\frac{d^D\!p}{(2\pi\hbar)^{D}}~e^{\frac{i}{\hbar}p(x - x')}A\left(\frac{x + x'}{2}, p\right),
\end{eqnarray}  
which differs from (\ref{eq:matrixfourier}) by the projection operator, $\Pi$.

Since ${\cal S}$ and ${\cal S}^{-1}$ are the inverse map to each other, the algebraic manipulations in the previous subsection 
for ${\cal O}$ are performed in terms of ${\cal S}$ without ambiguity. 
For example the product of two operators which are images 
of two functions by ${\cal S}$ is evaluated as
\begin{eqnarray}
       {\cal S}^{-1}({\cal S}(A){\cal S}(B))(\zeta) =     e^{\frac{i\hbar}{2}\nabla_{z}J_0\nabla_{z'}}A(z)B(z')|_{z = z'= \zeta},\label{spstar}
\end{eqnarray}
where
\begin{eqnarray}
	J_0 = \left( \begin{array}{cc} 0 & 1\\
          -1 & 0  \end{array} \right) .
\end{eqnarray}
Let us denote r.h.s. of eq.(\ref{spstar}) by $A\bullet B(\zeta)$. Thus, ${\cal S}(A){\cal S}(B) = {\cal S}(A\bullet B)$.

In order to implement the framework of the quantum theory we must introduce a 
notion of states. In the Moyal quantization the corresponding entity is the Wigner function defined by
\begin{eqnarray}
       f_W(\zeta) = {\cal S}^{-1}(|\psi \rangle \langle \psi|),
\end{eqnarray}
where $|\psi \rangle$ is the state vector in the usual quantum theory.
The Wigner function is a real function of the phasespace, satisfying the
pure state condition
\begin{eqnarray}
       f_W\bullet f_W = f_W.  \label{pure}
\end{eqnarray}
The time independent {Schr\"{o}dinger} equation $\hat{H}|\psi\rangle = E|\psi\rangle$ is written in 
terms of the Wigner function as
\begin{eqnarray}
       {\cal S}^{-1}(\hat{H})\bullet f_W = Ef_W.  \label{spstargen}
\end{eqnarray}
In this equation  information about the constraints is involved only in ${\cal S}^{-1}(\hat{H})$.

Now let us rewrite the above formulas  in terms of ${\cal O}$.
Note that the relations between ${\cal S}^{-1}$ and ${\cal O}$ are written, in an obvious notation, as
\begin{eqnarray}
           {\cal S}^{-1}{\cal O} &=& \tilde{\Pi}, \label{hih}\\
          \langle x|{\cal O}{\cal S}^{-1}(\hat{A})|y\rangle &=& \delta(n\cdot (x - y))\int d(n\cdot u)\left<\frac{1}{2} n\cdot u  + \Pi x\Biggm|\hat{A}\Biggm|-\frac{1}{2} n\cdot u  + \Pi y\right>,\nonumber  \\
\end{eqnarray}
where $n$ is a zero vector of $\Pi$, and
\begin{eqnarray}
 \tilde{\Pi} A(q, p) = A(\Pi q, \Pi p).
\end{eqnarray}
In the case with no constraint, where $\Pi = 1$ and $n = 0$, the above relations show that ${\cal S}^{-1}$ is just  the inverse of ${\cal O}$. For the existence of constraint we see from (\ref{hih}) that
\begin{eqnarray}
           {\cal O} = {\cal S} \tilde{\Pi}.\label{htos}
\end{eqnarray}
The two kinds of products of functions, defined above, are related by
\begin{eqnarray}
           \tilde{\Pi}(A\bullet B) = \tilde{\Pi}(A\star B).
\end{eqnarray}

The reason why multiple functions correspond to a single operator by the ${\cal O}$ map is 
in that $\tilde{\Pi}$ is a singular map, 
 a direct consequence of the existence of the constraint, $n\cdot\hat{z} = 0$.
(This is also seen in the definition (\ref{eq:d_invfourier}), where  one can freely add the term proportional 
to $n\cdot\hat{z}$ in the exponent in eq.(\ref{eq:oprinvfourier}), causing the ambiguity of the 
corresponding function.)
 If one restricts the function space to those functions satisfying $\tilde{\Pi} f = f$, the two maps, ${\cal O}$ 
and ${\cal S}$, coincide, and ${\cal O}$ is a one-to-one mapping on it. Let us denote the subspace of whole 
function space as ${\cal C}$:
\begin{eqnarray}
        {\cal C} = \{f | \tilde{\Pi} f = f\}.
\end{eqnarray}
If one restrict the ${\cal O}$-map to ${\cal C}$, then it is a one-to-one map, and we can write ${\cal S}^{-1}$ as 
${\cal O}^{-1}$ on ${\cal C}$.
The {Schr\"{o}dinger} equation (\ref{spstargen}) for $f_W \in {\cal C}$ becomes
\begin{eqnarray}
       {\cal O}^{-1}(\hat{H})\star f_W = Ef_W,  \label{stargen}
\end{eqnarray}
with the pure state condition
\begin{eqnarray}
       f_W\star f_W = f_W.  \label{hpure}
\end{eqnarray}
In these forms information about the constraint is involved both in $\hat{H}$ {\it and} 
the definition of the $\star$ product.

We see the ${\cal O}$ form of the {Schr\"{o}dinger} equation requires the state function $f_W$ to satisfy 
$\tilde{\Pi} f_W = f_W$, {\it i.e.},the Wigner function should be 
defined  essentially on the constraint surface. In the next section we 
will see the existence of more natural formulation, where the constraint is imposed in a 
geometrical way at the outset and the entire quantum equations are defined on the constraint 
surface.

\section{Reconsideration of the Dirac Quantization}

A quantum system with the second class constraints, such as the gauge fixed 
version of a quantum gauge theory, might be treated in the 
Dirac method. A sequence of canonical transformations are used to 
rearrange  the redundant variables to form a set of canonical 
pairs equivalent to the  constraints.  The commutators are replaced
by their Dirac brackets so that the redundant variables are 
consistently set to zero. This procedure is, however,  a 
cumbersome one for some cases, and the more convenient quantization schemes 
 are thought to be favorable, especially in the gauge theories.

In a sense the second class constraint  determines a geometry of the space spanned
by the dynamical variables of a classical theory. In order to quantize the system 
one might promote the geometry (described by c-numbers) to a {\it quantum} geometry 
(on q-numbers), and it may require some challenging attempts (one of which is that by Dirac).
In this account the Moyal quantization seems to be a suitable framework of 
the theories with the second class constraints, since it uses only c number functions, 
and a geometrical notion is easily introduced.  

In the previous section we have considered the Moyal quantization of the 
theories with the second class constraint, starting with the quantum theories 
of the Dirac scheme.
Conversely, let us attempt to construct a Moyal quantization in which the Dirac 
theory is {\it derived} as a consequence.
We propose the following correspondence between a quantum operator and a 
function on the phasespace:
\begin{eqnarray}
     \hat{A} = {\cal F}(A) \equiv  \int \frac{du d\zeta}{(2\pi)^{2D}}~e^{iu(\hat{z}  - \zeta)}A(\zeta)\delta(\Phi_a(\zeta))\delta(\Phi_a(u)),\label{kk}
\end{eqnarray}
where $\Phi_a(\zeta) = 0$ are second class constraints, and $\hat{z}_i = (\hat{x}_i, \hat{p}_i)$ satisfy the 
{\it canonical} CR;
\begin{eqnarray}
    {[}\hat{x}_i, \hat{p}_j{]} _{opr}= i\hbar\delta_{ij}.
\end{eqnarray}
Note that the range of integration in (\ref{kk}) is restricted to the constraint surface by the delta functions, while the operators $\hat{z}_i$  are {\it not} subject to any constraint.
Though we have written in (\ref{kk}) the formula for general (2nd class) constraint for the sake of future investigation, we will consider in what follows the special case, where  $\Phi(\zeta) = n\cdot\zeta$.
In this case the 
calculation is easy and some essential ingredients of the proposal are displayed.

In the coordinate system in which the normal vector $n$ to the constraint surface is given by $n = \tilde{n} = (1, 0,....,0)$, we see from (\ref{kk}) that
\begin{eqnarray}
       {\cal F}(\zeta_i) = \left\{ \begin{array}{c} \hat{z}_i \qquad i \ne 1\\
                                  0 \qquad i = 1 ,\end{array}   \right. 
\end{eqnarray}
{\it i.e.}, ${\cal F}(\zeta) = \tilde{n}\cdot\hat{z}$.
For general $n$, one rotates all variables in (\ref{kk}) by
\begin{eqnarray}
 n_i = R_{ij}\tilde{n}_j,   \qquad   \tilde{z}_j = \hat{z}_iR_{ij},\qquad  \tilde{u}_j = u_iR_{ij}, \qquad \tilde{\zeta}_j = \zeta_iR_{ij},
\end{eqnarray}
\begin{eqnarray}
\tilde{\Pi}_{ij} = \delta_{ij} - \frac{\tilde{n}_i\tilde{n}_j}{\tilde{n}^2},
\end{eqnarray}
where $R \in SO(D)\times SO(D)$.
Then we have
\begin{eqnarray}
{\cal F}(\zeta_i) &=&  R_{ii'}\int d\tilde{u}d\tilde{\zeta}~e^{i\tilde{u}(\tilde{z}  - \tilde{\zeta})}\tilde{\zeta}_{i'}\delta(\tilde{n}\cdot \tilde{\zeta})\delta(\tilde{n}\cdot\tilde{u})\nonumber  \\
 &=& R_{ii'}\tilde{\Pi}_{i'j'}\tilde{z}_{j'}\nonumber  \\
 &=& \Pi_{ij}\hat{z}_j. \label{fofz}
\end{eqnarray}
This means that ${\cal F}$ is a map from functions of whole phasespace to the space of 
operators which satisfy $n\cdot\hat{z} = 0$.
(Note that ${\cal F}(n\cdot\zeta) = 0$, but this does {\it not} mean $n\cdot\zeta = 0$.)
The Dirac quantization scheme is reproduced by ${\cal F}$, {\it i.e.},
\begin{eqnarray}
        {[}{\cal F}(q_i), {\cal F}(p_j){]}_{opr} = i\hbar\Pi_{ij},
\end{eqnarray}
where r.h.s. coincides with the Dirac bracket $\{q_i, p_j\}_D$ (times $\hbar$) calculated from our constraint.
The product  of two operators which are images by ${\cal F}$ reads  
\begin{eqnarray}
       {\cal F}(A){\cal F}(B) &=& {\cal F}(C),
\end{eqnarray}
\begin{eqnarray}
        C &=& \int dv_Adv_B~e^{\frac{2i}{\hbar}v_AJ_0v_B}A(\zeta + v_A)B(\zeta + v_B)\delta(n\cdot v_A)\delta(n\cdot v_B).\label{fstar}
\end{eqnarray}
By using these equations one can derive the product rule defined by ${\cal F}(A){\cal F}(B) = {\cal F}(A\star B)$. 
It turns out that the $\star$ product thus defined is the same as (\ref{eq:starproduct}) 
in the case of the ${\cal O}$-map.
In fact, ${\cal O}$ and ${\cal F}$ are the same map in the case of the flat constraint condition, $n\cdot \zeta = 0$.
This is explicitly seen from the fact
\begin{eqnarray}
      {\cal S}^{-1}{\cal F} = \tilde{\Pi},
\end{eqnarray}
and (\ref{htos}).
At present, however, it is an open problem whether in general cases the equivalence of  ${\cal O}$ and ${\cal F}$ holds or not.

To proceed further we must introduce a notion of states which could be pure or mixed ones.
It seems natural to adopt a real function of the phasespace  as the state vector. Furthermore we 
assume that the function $f$ satisfies $\tilde{\Pi} f = f$, {\it i.e.}, $f \in {\cal C}$.
We also assume the pure state condition for $f$, defined by (\ref{hpure}), which means ${\cal F}(f)$ is a 
projection operator, denoted by  $|\psi \rangle \langle \psi|$, establishing the correspondence $f \leftrightarrow |\psi\rangle$.

The Schr\"{o}dinger equation in the Dirac quantization is written in our scheme as
\begin{eqnarray}
         i \hbar \frac{\partial}{\partial t}f = H \star f. \label{eqom}
\end{eqnarray}
These equations might be solved for the Wigner functions which belong to ${\cal C}$ and satisfy the pure state condition.

In concluding this section we note a difference between ${\cal O}$ and ${\cal F}$ both in their principles and  
in practical manipulations.
The ${\cal O}$-formulation is equivalent to the Dirac scheme as a quantization procedure, since the former 
employs the quantization condition by Dirac. On the other hand the ${\cal F}$-formulation incorporates the
constraint condition in the operator function correspondence relation, while the quantization condition 
in this case is the canonical one without any constraint, and the CR of Dirac is {\it derived} as a 
consequence.
In the practical aspect, the definition of star-product of two functions in the case of ${\cal O}$ is 
expressed in a manifest form by use of the 
Campbel-Hausdorff formula which gives an expression of product of two exponentiated operators, $e^A$ 
and $e^B$.
In the case with general constraints it is extremely difficult to write down the above formula in a 
closed form.
On the contrary the product rule in terms of ${\cal F}$ is written in the form (\ref{fstar}) for general cases 
leaving only integration containing some delta functions.

\section{Applications}

\subsection{The harmonic oscillator with a constraint}

In this subsection we discuss how the Moyal quantization is subject to modification when there exist constraints by exemplifying a harmonic oscillator with a constraint. Before we get into discussion, we first show the Dirac brackets for the constrained system and its quantization, upon which our discussion is based. 
Let $n$ be a D-dimensional constant vector, and we consider a harmonic oscillator on the surface normal to the the vector $n$ . The classical Hamiltonian is given by $H(p,q)=(q^2+p^2)/2$. The coordinate $x \in R^D$ of arbitrary point on the surface satisfies the primary constraint
\begin{equation}
\Phi_1=n \cdot x \approx 0.
\end{equation}
The secondary constraint is derived from $\dot \Phi_1=0$ as
\begin{equation}
\Phi_2=n \cdot p \approx 0.
\end{equation}
Then, the Dirac brackets are given by
\begin{eqnarray}
\{q_i,q_j\}_D&=&0,\\
\{p_i,p_j\}_D&=&0,\\
\{q_i,p_j\}_D&=&\Pi_{ij},
\end{eqnarray}
where $\Pi_{ij}$ is given by
\begin{equation}
\Pi_{ij}=\delta_{ij}-\frac{n_i n_j}{n^2}.
\end{equation}
Here we note that the matrix $\Pi$ is a projection satisfying $\Pi \Pi=\Pi$, and has a singular property $det \Pi=0$.
Their quantization is obtained by replacing the Dirac brackets 
by the operator CR as
\begin{eqnarray}
{[}{\hat q}_i, {\hat q}_j {]}_{opr} &=& 0,\\
{[}{\hat p}_i, {\hat p}_j {]}_{opr} &=& 0,\\
{[}{\hat q}_i, {\hat p}_j {]}_{opr} &=& i \hbar \Pi_{ij}.
\end{eqnarray}
In order to get a classical hamiltonian for the Moyal quantization, we start with the classical hamiltonian defined above, and then use the ${\cal S}^{-1}$ map followed by ${\cal O}$ map successively to obtain
\begin{eqnarray}
H_W &=& h^{-D}\int dq~e^{-iu \cdot v/\hbar}\langle q+\frac{\xi}{2}|{\cal O}(H(q,p))|q-\frac{\xi}{2}\rangle \nonumber \\
&=&\frac{1}{2}(x \cdot \Pi x+p \cdot \Pi p).
\end{eqnarray}
By using the Hamiltonian $H_W$, the the {Schr\"{o}dinger} equation or the $\star$-gen value equation discussed in section 2.2 reads
\begin{equation}
H_W\star f(x,p)=Ef(x,p),
\label{eq:stareq}
\end{equation}
with the condition restricting the function space:
\begin{equation}
\tilde \Pi f(x,p)=f(x,p).
\label{eq:picond}
\end{equation}
Our strategy is to solve the {Schr\"{o}dinger} equation to obtain $f(x,p)$, while checking if $f(x,p)=f(\Pi x,\Pi p)$ is satisfied.

There are two ways of solving this equation, an analytic method and an algebraic method. The former solution is the 1st quantization-like method, and the latter solution is the 2nd quantization-like method.

\subsubsection{Analytic method}

The real part of Eq.(\ref{eq:stareq}) is given by
\begin{equation}
\Big\{ \frac{1}{2}\Big( x \cdot \Pi x+p \cdot \Pi p \Big)-
\frac{1}{2}\Big(\frac{\hbar}{2}\Big)^2 \Pi_{ij}
\Big(\frac{\partial ^2}{\partial p_i \partial p_j}+\frac{\partial ^2}{\partial x_i \partial x_j}\Big)-E \Big\}f(x,p)=0,
\label{eq:realstareq}
\end{equation}
and the imaginary part of Eq.(\ref{eq:stareq}) by
\begin{equation}
(x \cdot \Pi \frac{\partial}{\partial p}-p \cdot \Pi \frac{\partial}{\partial x})
f(x,p)=0.
\label{eq:imgstareq}
\end{equation}
By taking the condition (\ref{eq:picond}) into account, the imaginary part equation is solved by $f=f(z)$ with $z=4H_W$. Then (\ref{eq:realstareq}) is rewritten by using $z$ as
\begin{equation}
\frac{z}{4}f-\hbar^2(z f''+(D-1)f')-Ef=0,
\end{equation}
where $f'=\partial f/\partial z$ and $f''=\partial^2 f/\partial z^2$.
By introducing $\zeta$ by
\begin{equation}
\zeta=\frac{z}{\hbar},
\end{equation}
the equation is written as 
\begin{equation}
\frac{\zeta}{4}f-\{\zeta f''+(D-1)f'\}-Ef=0,
\end{equation}
where the derivatives are taken with respect to $\zeta$.
Setting 
\begin{equation}
f(\zeta)={\rm exp}\Big(-\frac{\zeta}{2}\Big)L(\zeta),
\end{equation}
we obtain
\begin{equation}
\zeta L''+(D-1-\zeta)L'+\Big(\frac{E}{\hbar}-\frac{D-1}{2}\Big)L=0.
\end{equation}
The solution is given by the Laguerre polynomial:
\begin{equation}
L_n^{(D-2)}(\frac{z}{\hbar}),
\end{equation}
for $E_n=(n+(D-1)/2)\hbar$, ($n=0,1,2,\cdots$). The eigenvalue shows the reduction of the dimension by 1 which is reasonable judging from the present constraint.

\subsubsection{Algebraic method }

We can also solve Eq.(\ref{eq:stareq}) algebraically, besides the analytic method in the previous subsection. This algebraic method is analogous to the solution by using the creation and annihilation operators.
Define $a_k$ and $a_k^\dagger$ by
\begin{eqnarray}
a_k&=&\frac{1}{\sqrt 2}(x_k+ip_k),\\
a_k^\dagger&=&\frac{1}{\sqrt 2}(x_k-ip_k).
\end{eqnarray}
Then their CR read
\begin{eqnarray}
&&[a_k,a_l]=[a_k^\dagger,a_l^\dagger]=0,\\
&&[a_k,a_l^\dagger]=\hbar \Pi_{kl}.
\end{eqnarray}
The hamiltonian $H_W$ is written by use of $a_k$ and $a_l$ as
\begin{eqnarray}
H_W&=&\frac{1}{2}(x \cdot \Pi x+p \cdot \Pi p) \cr
   &=&\frac{1}{2}\Pi_{kl}(a_k \star a_l^{\dagger}+a_k^{\dagger} \star a_l)\cr
   &=&\Pi_{kl}a_k^{\dagger} \star a_l+\frac{1}{2}(D-1)\hbar.
\end{eqnarray}
Define the vacuum $f_0(x,p)$ by
\begin{equation}
\Pi_{kl}a_l \star f_0(x,p)=f_0(x,p) \star \Pi_{kl}a_l^\dagger=0, ~~{\rm for} ^\forall k.
\end{equation}
with the normalization condition
\begin{equation}
f_0(x,p) \star f_0(x,p)=f_0(x,p).
\end{equation}
We define the n-th state $f_n(x,p)$ by
\begin{eqnarray}
&&f_n(x,p)\cr
=&&\frac{1}{n!}\Big(\frac{1}{\hbar}\Big)^n \Pi_{i_1,j_1}\Pi_{i_2,j_2} \cdots
\Pi_{i_n,j_n}a_{i_1}^{\dagger} \star a_{i_2}^{\dagger} \star \cdots 
\star a_{i_n}^{\dagger}\star f_0 \star a_{j_n} \star a_{j_{n-1}} \cdots
\star a_{j_1}, \nonumber \\
\end{eqnarray}
which satisfy the normalization conditions
\begin{equation}
f_n(x,p) \star f_n(x,p)=f_n(x,p).
\end{equation}
We can then show that
\begin{equation}
H_W \star f_n=E_n f_n,
\end{equation}
where $E_n=(n+(D-1)/2)\hbar$. This is identical to the result obtained in the previous subsection.

The explicit form of the vacuum function $f_0(x,p)$ is given by the Weyl vacuum
\begin{equation}
f_0(x,p)={\rm exp}\Big(-\frac{2 H_W}{\hbar} \Big).
\end{equation}
Note that they are different from the vacuum for the non-constrained case, because
\begin{eqnarray}
a_k \star f_0(x,p) \ne 0,\\
f_0(x,p) \star a_k^\dagger \ne 0.
\end{eqnarray}
In order to see for which operators they play the role of vacuum, 
it is useful to introduce $\alpha_k$ and $\alpha_k^\dagger$ by
\begin{eqnarray}
\alpha_k=\Pi_{kl}a_l,\\
\alpha_k^\dagger=\Pi_{kl}a_l^\dagger,
\end{eqnarray}
which satisfy $n \cdot \alpha=0$ and $n \cdot \alpha^{\dagger}=0$. 
We can show that the vacuum $f_0(x,p)$ is annihilated by $\alpha_k$ when operated from the left and by $\alpha_k^\dagger$ from the right,
\begin{equation}
\alpha_k \star f_0(x,p)=f_0(x,p) \star \alpha_k^\dagger=0, ~~{\rm for} ^\forall k.
\end{equation}
By using $\alpha_k$ and $\alpha_k^\dagger$, the Hamiltonian $H_W$ and the $\star$-gen functions are rewritten in a simpler way. The hamiltonian $H_W$ reads
\begin{equation}
H_W=\frac{1}{2}(\alpha_k \star \alpha_k^\dagger+\alpha_k^\dagger \star \alpha_k).
\end{equation}
By using the vacuum, the n-th state is rewritten as
\begin{equation}
f_n(x,p)=\frac{1}{n!}\Big(\frac{1}{\hbar}\Big)^n
(\alpha_k ^\dagger \star)^n f_0 (\star \alpha_k)^n.
\end{equation}
Here we note that these states satisfy (\ref{eq:picond}), since they are constructed by $\alpha_k$ and $\alpha_k^\dagger$.

As for the vacuum, we can also define it in an unusual way.\cite{Koi} In order to give the explicit form of the 2nd type of vacuum, we need to introduce the normal $\star$ ordering denoted by the double dots, which put all $\alpha_k$ to the right of all $\alpha_k^\dagger$. The 2nd type of vacuum is given by
\begin{eqnarray}
\tilde f_0(x,p)&=&:{\rm exp}\Big(-\frac{\alpha_k ^\dagger \star \alpha_k}{\hbar} \Big): 
\cr
&=&1-\frac{1}{\hbar}\alpha_k ^\dagger \star \alpha_k+\frac{1}{2}
\alpha_k ^\dagger \star \alpha_l ^\dagger \star \alpha_k \star \alpha_l+\cdots.
\end{eqnarray}
By using this vacuum, we can also construct the n-th state as well, and as is clear from the construction, these states satisfy (\ref{eq:picond}).

\subsection{Electro-magnetism in the Coulomb gauge}

We consider here the electro-magnetic dynamics in the Moyal scheme as another example of constrained system.
In the gauge invariant theories, there exist the 1st class constraints 
associated with the gauge invariance.
The treatment of such a constrained system is
well-understood in classical and ordinary quantum field theories.
The procedure described in section 2 and 3 is applicable also to the field theory simply by replacing the discrete indices by a continuous variable.

Since the Moyal quantization is deeply related to the canonical structure 
of the theory, we give here a rather detailed description of the 
canonical theory.
We start with the classical Lagrangian  $L$ of 
the electro-magnetic dynamics with external conserved current $J^{\mu}$:
\begin{eqnarray}
L = \int d^3x\left(-\frac{1}{4} F_{\mu\nu}F^{\mu\nu}-eA_{\mu}J^{\mu}\right),
\label{em:lag}
\end{eqnarray}
where $e$ is the electric charge, 
$A_{\mu}$ is the vector potential, and  
$F_{\mu \nu}:=\partial_{\mu}A_{\nu}-\partial_{\nu}A_{\mu}$ is the electro-magnetic field strength.

The action  is invariant under the gauge transformation:
\begin{eqnarray}
A_{\mu}(x)\longrightarrow A_{\mu}(x)+\partial_{\mu} \theta  (x),
\label{em:gaugetrans}
\end{eqnarray}
where $\theta(x)$ is an arbitrary function of $x$. 
The conjugate momenta of $A_{\mu}$ are defined by the functional derivative of 
$L$ with respect to $\dot A_{\mu}$:
\begin{eqnarray}
\pi^{\mu}(x) = \frac{\delta L}{\delta \dot A_{\mu}(x)}.
\end{eqnarray}
We obtain $\pi^i=F^{0i}:=E^i$. But, as $\dot A^0$ is not included in Lagrangian (\ref{em:lag}), the 0-th component of the above equation yields the primary constraint 
\begin{eqnarray}
\pi_{0}:=\phi_1 \approx 0.
\end{eqnarray}

The total Hamiltonian of the system is given by
\begin{eqnarray}
H = \int d^3x \left(\frac{1}{2}\pi^i \pi_i + \frac{1}{2}B_iB^i - eA_iJ^i 
+ A_0(\partial_i \pi^i - eJ^0) - \lambda_0 \pi^0+\partial_i(\pi^i A_0)\right),
\cr
\label{em:ham1}
\end{eqnarray}
where $B_i:=\varepsilon^{ijk} \partial_j A_k $ is the magnetic field 
and $\lambda_0$ is the Lagrange multiplier.
We can also obtain the secondary constraint from the condition for consistency with the time evolution
 of $\pi_0$ :
\begin{eqnarray}
 \dot \pi_0 =\{\pi_0, H\}_P+\lambda \{ \pi_0 ,\pi_0 \}_P=\partial_i \pi_i - eJ_0:=\phi_2\approx 0.
\label{em:constf2}
\end{eqnarray}
Note that the Poisson bracket $\{ , \}_P$ is defined by functional derivatives.
This equation is known as Gauss's law and 
appear in the fourth term of (\ref{em:ham1}).
This means that $A_0$ can be absorbed into the Langange multiplier (say $\lambda_2$)
for the constraint $\phi_2$.
The above two constraints are the 1st class, generating the gauge 
transformations corresponding to (\ref{em:gaugetrans}) in the Lagrangian 
formulation.
For a functional $F[A, \pi]$ of the canonical variables, 
the gauge transformation in the canonical theory is expressed as $\delta F = \{F, \int \epsilon_a\Phi_a \}_P$.

We choose the Coulomb gauge as the gauge fixing condition:
\begin{eqnarray}
\chi_1 =  \partial_iA_i \approx 0.
\label{em:constc1}
\end{eqnarray}
The preservation of $\chi_1 = 0$ for the time evolution requires $\partial_i(\pi_i-\partial_iA_0) \approx 0$, which together with Gauss's law (\ref{em:constf2}) gives Poisson's equation 
\begin{eqnarray}
\chi_2 = \Delta A_0 - eJ_0 \approx 0.
\end{eqnarray}
Then these constraints $\Phi_a:=(\phi_1$,$\phi_2$,$\chi_1$,$\chi_2)$ satisfy 
$\det \{\Phi_a,\Phi_b\}_P \ne 0$, 
and constitute the 2nd class constraints as a whole.
Therefore the classical equations of motion are given by variation of the modified Hamiltonian $H_*$ 
with respect to independent variables $(A_i, \pi_i), (\lambda_a, \Phi_a)$:
\begin{eqnarray}
H_* = \int d^3x \left(\frac{1}{2}\pi^i \pi_i + \frac{1}{2}A_i\Delta A_i - eA_i J^i
-\lambda_a \Phi_a\right).
\label{em:ham2}
\end{eqnarray}
Since the constraints $\Phi_a$ are 2nd class, 
we compute the Dirac bracket $\{ , \}_D$ of these variables
\begin{eqnarray}
\{A_i(x), \pi_j(x')\}_D &\approx & \Pi_{ij}(x - x'),\\
\{A_i(x), A_j(x')\}_D &\approx& \{\pi_i(x), \pi_j(x')\}_D \approx 0,
\end{eqnarray}
where
\begin{eqnarray}
\Pi_{ij}(x - x') &=& \left(\delta_{ij} - \frac{\partial_i\partial_j}{\Delta}\right)\delta^3(x-x').
\end{eqnarray}
The functional matrix $\Pi$ is a projection as is in the example of the 
previous subsection. Here and hereafter we restrict our attention  only to $A_i$ and $\pi_i$ and omit $A_0$ and $\pi_0$, since these variables do not play the essential part in the discussion.

The quantization is implemented by replacing the Dirac brackets by the operator CR as
\begin{eqnarray}
{[}\hat{A}_i(x), \hat{\pi}_j(x'){]} &=& i\hbar\Pi_{ij}(x - x'),\\
{[}\hat{A}_i(x),\hat{A}_j(x'){]} &=& {[}\hat{\pi}_i(x),\hat{\pi}_j(x'){]} = 0.
\end{eqnarray}
A representation of these CRs is defined on the functional space of functions 
$s_i \in C^{\infty }(R^3)$ as
\begin{eqnarray}
      \hat{A}_i(x) &=& \int dx' \Pi_{ij}(x - x')s_i(x'),\label{qqq1}\\
      \hat{\pi}_i(x) &=& -i\hbar \int dx' \Pi_{ij}(x - x')\frac{\delta}{\delta s_j(x')}  + e\frac{\partial_i}{\Delta }J_0(x),\label{qqq2}\\
     |a\rangle &=& \prod_{x i}\delta(s_i(x) - a_i(x)),\label{qqq3}
\end{eqnarray}
where the second term on the r.h.s. of (\ref{qqq2}) comes from the constraint for 
the momentum operator $\hat{\pi}_i$.
The state vector $|a\rangle$ is the eigenstate of $\hat{A}_i(x)$:
\begin{eqnarray}
     \hat{A}_i(x)|a\rangle = \int dx' \Pi_{ij}(x - x')a_i(x')|a\rangle.
\end{eqnarray}

Now the WW correspondence is introduced as follows.
Given a functional $F$ of $A_i$ and $\pi_i$, the image of ${\cal O}$-map 
is defined by
\begin{eqnarray}
{\cal O}(F) =  F\left[-i\frac{\delta}{\delta u}, -i\frac{\delta}{\delta w}\right]e^{\int d^3x~ i(u\hat{A} + w\hat{\pi})} \Biggm|_{u=w=0}.\label{em:heart}
\end{eqnarray}
Conversely, the ${\cal S}^{-1}$ image of a given operator $\hat{F}$ is 
defined by
\begin{eqnarray}
{\cal S}^{-1}(\hat{F}) = \int (D\psi)~e^{-\frac{i}{\hbar}\int d^3x~\pi\cdot\psi}\left<A + \frac{\psi}{2}\biggm|\hat{F}\biggm|A - \frac{\psi}{2}\right>,\label{em:spinv}
\end{eqnarray} 
where the inner product represents the functional integration over all functions $s_i(x)$. The star product of two functionals, $F$ and $G$, is given by
\begin{eqnarray}
&&(F\star G)[A,\pi] \nonumber\\
&=&   e^{\frac{i\hbar}{2} \int dxdx'\Pi_{ij}(x - x')\left(\frac{\delta}{\delta A_{i}(x)} \frac{\delta}{\delta \pi'_{j}(x')} - \frac{\delta}{\delta \pi_{i}(x)}\frac{\delta}{\delta A'_{j}(x')}\right)} 
F[A,\pi]G[A',\pi'] \biggm|_{A=A',\pi=\pi'},
\end{eqnarray}
by which the Moyal bracket for these functionals is defined by
\begin{eqnarray}
\{F, G\}_M[A,\pi] = (F\star G - G\star F)[A,\pi].
\end{eqnarray}
 
The Weyl symbol of the Hamiltonian is
\begin{eqnarray}
    H_W = {\cal S}^{-1}(\hat{H}) = \int d^3xd^3x'
\left(\frac12\pi_i(x)\Pi_{ij}(x - x')\pi_j(x') 
+ \frac12\tilde{A}_i(x)\Pi_{ij}(x - x')\tilde{A}_j(x')\right),\cr
\end{eqnarray}
where
\begin{eqnarray}
  \tilde{A}_i(x) = A_i(x) - eJ_i(x).
\end{eqnarray}
The Schr\"{o}dinger equation is
\begin{eqnarray}
     i \hbar \frac{\partial}{\partial t}F = H_W \star F,\label{em:schr}
\end{eqnarray}
with the pure state condition
\begin{eqnarray}
      F\star F = F.
\end{eqnarray}
$F$ is the Wigner functional having all the information of the 
electro-magnetic field.

In this subsection, we have shown the Moyal quantization of the electro-magnetism with the conserved external current as one of the examples of the constrained system. Eq.(\ref{em:schr}) is the basic equation of the present formulation, by which 
all the physical quantities might be  calculated.

\section{Summary and Discussion}

We have studied the Moyal quantization when there exist constraints. In the quantization of the classical system with the 2nd class constraints, it is known that the Dirac bracket is to be used instead of the Poisson bracket in transferring to to the quantum theory. The Weyl correspondence relating a quantum operator to a function called the Weyl symbol is subject to modification when there exist constraints. In order to study this, we assume a constraint limiting the phasespace to the surface normal to a constant vector. To summarize the process obtaining the Weyl symbol, we start with classical fields, get the quantum fields from them, and finally the Weyl symbols from the quantum fields. 

When there is no constraint, the classical fields and the Weyl symbols are identical. This is assured because the WW correspondence is a one-to-one map. However, when there are constraints, we find that the classical fields and the Weyl symbols are not necessarily identical. To show this situation by an explicit example, we adopt one of the simplest constraints mentioned above. 

We showed two formulations of the Moyal quantization for the constrained system. One is the well-known Weyl correspondence for the non-constrained system, which uses the operators $\hat z$ in the definition. We call it the ${\cal O}$ formulation. The other uses the state instead of the operators $\hat z$ and  we call this the ${\cal S}$ formulation. In contrast to the Weyl correspondence for the non-constrained case, the ${\cal O}$ formulation has a trouble that the correspondence is not a one-to-one map. This is evaded in the ${\cal S}$ formulation. Another difference between the two formulations also appears in the definition of the product between two Weyl symbols i.e. the $\star$ product. In the ${\cal O}$ formulation, the product reflects the constraint, and so its form is different from that for the non-constrained case. The product in the ${\cal S}$ formulation is the same star product as that for the non-constrained case. This difference comes from the fact that we use the operator $\hat z$, which constitute the non canonical CRs, in the definition of the Weyl correspondence in the ${\cal O}$ formulation. Then one may wonder what is the effect of the constraints in the ${\cal S}$ formulation. All information about the constraints can be found in the Weyl symbols obtained from operators by the inverse map of the ${\cal S}$ there.

Now we summarize the procedure. In the Moyal quantization for the constrained system, we start with a classical hamiltonian with constraints, and the Dirac brackets are used for the quantization. The ${\cal O}$ map is used to obtain the quantum Hamiltonian from the classical one. Finally, by using ${\cal S}^{-1}$ map, we obtain a Hamiltonian in terms of the Weyl symbols. Though there can be two kinds of the star products for the ${\cal O}$ and ${\cal S}$ formulations, the so-called $\star$-gen equation derived from the two formulations becomes identical, and so it does not make difference for a practical use.

In section 3, we proposed a Weyl correspondence for the constrained case, which impose restrictions in the phasespace and its dual space integration by introducing the $\delta$ functions of the 2nd class constraints.  The coordinate and momentum operators are assumed to satisfy the canonical CRs and so we do not use the Dirac bracket method.  We only restrict the integration space. As the result, we obtain an intriguing result that the same CRs as those derived from the Dirac brackets emerge naturally. It is immature to conclude that the present formulation is equivalent to the Dirac bracket formulation, and so further survey is necessary.
Although the 2nd class constraints, which are expressed as functions in the phasespace, should have a clear geometrical meaning, their geometrical meaning is not so clear in the Dirac bracket quantization. On the other hand, the present Moyal quantization for the constrained case is formulated by incorporating the constraints, which are c-number functions, as the $\delta$ functions in the phasespace and its dual space. Therefore, it is possible to understand them geometrically. 

Our constraint, which gives constant $g_{ij}(q,p)$ in Eq.(\ref{eq:gcr}), is one of the various 2nd class constraints.  For a general $g_{ij}(q,p)$, the Moyal quantization in terms of the ${\cal O}$ formulation would become too complicated, and it seems difficult to pursue further. However, when there is a simplification like symmetries, the formulation in section 3 would be useful and would shed light on these quantization.


\newpage
\begin{thebibliography}{99}
\bibitem{Moy}J.~E.~Moyal, Proc.~Cambridge Phil.~Soc.~{\bf 45}~(1949),~90.
\bibitem{OsMo}T.~A.~Osborn and F.~H.~Molzahn,~Annals Phys.,~{\bf 241}~(1995),~79
\bibitem{Dirac}P.~A.~M.~Dirac,~Can.~J.~Math.,~{\bf 2}~(1950),~129	
\bibitem{Koi}T.~Koikawa, Prog.~Theor.~Phys.~{\bf 106}~(2001),~1027;{\bf 107}~(2002),~1061.
\end{thebibliography}
\end{document}